\documentclass[10pt,conference]{IEEEtran}
\usepackage{caption}
\usepackage{cite}
\usepackage{mathtools}
\usepackage{amsmath,amssymb,amsfonts}
\usepackage{amsthm} 
\usepackage{commath}

\usepackage{graphicx}
\usepackage{subcaption}
\usepackage[utf8]{inputenc}
\usepackage[english]{babel} 
\usepackage[export]{adjustbox}
\usepackage{wrapfig}

\usepackage{textcomp}
\usepackage{float}
\usepackage[table]{xcolor}
\usepackage{comment}
\usepackage[ruled,vlined, linesnumbered]{algorithm2e}

\usepackage[printonlyused]{acronym}
\acrodef{3D}{three-dimensional}
\acrodef{3GPP}{3rd generation partnership project}
\acrodef{5G}{fifth generation}
\acrodef{6G}{sixth generation}
\acrodef{A2C}{actor-critic}
\acrodef{AoD}{angle of departure}
\acrodef{AoA}{angle of arrival}
\acrodef{ANN}{artificial neural networks}
\acrodef{AWGN}{additive white gaussian channel}
\acrodef{BS} {base station}
\acrodef{BSs} {base stations}
\acrodef{BTS}{bayesian-thompson sampling}
\acrodef{CAB}{contextual multi-armed bandit}
\acrodef{CSI}{channel state information}
\acrodef{DFT}{discrete Fourier transform}
\acrodef{DNN}{deep neural networks}
\acrodef{DRL}{deep reinforcement learning}
\acrodef{DQN}{deep Q-network}
\acrodef{ESN}{echo state network}
\acrodef{FML}{fast machine learning}
\acrodef{fsp}{free space path loss}
\acrodef{gNB}{5G NR base station}
\acrodef{gUCB}{greedy upper confidence bound}
\acrodef{ICONICAL}{IIoT CONnectivity for mechanICAL systems}
\acrodef{IID}{independent and identical distribution}
\acrodef{LoS}{line-of-sight}
\acrodef{MAB}{multi-armed bandit}
\acrodef{MIMO}{multiple input multiple output}
\acrodef{ML}{machine learning}
\acrodef{MSE}{mean squared error}
\acrodef{nLoS}{non-line of sight}
\acrodef{NN}{neural networks}
\acrodef{NR}{new radio}
\acrodef{OFDM}{orthogonal frequency division multiple access}
\acrodef{mmWave}{millimeter wave}
\acrodef{MDP}{markov decision process}
\acrodef{POMDP}{partially observable Markov decision process}
\acrodef{RBD}{receiver beam direction}
\acrodef{ReLU}{rectifier linear units}
\acrodef{RL}{reinforcement learning}
\acrodef{RF}{radio frequency}
\acrodef{RIM}{RSSI Information Matrix}
\acrodef{RSS}{received signal strength}
\acrodef{SGD}{stocastic gradient descent}
\acrodef{SNR}{signal-to-noise ratio}
\acrodef{SSB}{subframe structure block}
\acrodef{SS}{subframe structure}
\acrodef{SCS}{sub-carrier spacing}
\acrodef{TTU}{travel time unit}
\acrodef{UAV}{unmanned aerial vehicle}
\acrodef{UAVs}{Unmanned aerial vehicles}
\acrodef{UE}{user equipment}
\acrodef{UEs}{user equipments}
\acrodef{ULA}{uniform linear array}
\acrodef{UMa}{urban macro-cellular}
\acrodef{UPA}{uniform planar array}
\acrodef{V2X}{vehicle-to-everything}
\acrodef{QoS}{quality of service}

\title{DQN-based Beamforming for Uplink mmWave Cellular-Connected UAVs}
\author{
\IEEEauthorblockN{Praneeth~Susarla\IEEEauthorrefmark{1}, Bikshapathi~Gouda\IEEEauthorrefmark{1}, Yansha~Deng\IEEEauthorrefmark{2}, Markku Juntti\IEEEauthorrefmark{1}, Olli S\'{i}lven\IEEEauthorrefmark{1}, Antti~T{ö}lli\IEEEauthorrefmark{1}}

\IEEEauthorblockA{\IEEEauthorrefmark{1}University of Oulu, Finland}
\IEEEauthorblockA{\IEEEauthorrefmark{2}King's College London, United Kingdom}
email: \IEEEauthorrefmark{1}firstname.lastname@oulu.fi, \IEEEauthorrefmark{2}firstname.lastname@kcl.ac.uk
}
\begin{document}
%
%
%
%
%
%
\maketitle
\begin{abstract}
Unmanned aerial vehicles (UAVs) are the emerging vital components of millimeter wave (mmWave) wireless systems. Accurate beam alignment is essential for efficient beam based mmWave communications of UAVs with base stations (BSs). Conventional beam sweeping approaches often have large overhead due to the high mobility and autonomous operation of UAVs. Learning-based approaches greatly reduce the overhead by leveraging UAV data, like position to identify optimal beam directions. In this paper, we propose a reinforcement learning (RL)-based framework for UAV-BS beam alignment using deep Q-Network (DQN) in a  mmWave setting. We consider uplink communications where the UAV hovers around 5G new radio (NR) BS coverage area, with varying channel conditions. The proposed learning framework uses the location information to maximize data rate through the optimal beam-pairs efficiently, upon every communication request from UAV inside the multi-location environment. We compare our proposed framework against Multi-Armed Bandit (MAB) learning-based approach and the traditional exhaustive approach, respectively and also analyse the training performance of DQN-based beam alignment over different coverage area requirements and channel conditions. Our results show that the proposed DQN-based beam alignment converge faster and generic for different environmental conditions. The framework can also learn optimal beam alignment comparable to the exhaustive approach in an online manner under real-time conditions. 
\end{abstract}
%
%
%
%
\section{Introduction}
\ac{UAVs} are envisioned as the vital ingredients for future wireless systems using \ac{mmWave}. 
Especially, the deployment of cellular-enabled UAV-\ac{UE}s (\textit{hereafter addressed as UAVs}) adds unique features pertaining to high mobility and autonomous operations to a myriad of civil applications, such as traffic surveillance, mineral exploration, internet drone delivery systems, etc.\cite{cellular_UAV_3}. 
The \ac{mmWave} frequencies with \ac{MIMO} beamforming and \ac{LoS} dominant connectivity enable high-speed data access for \ac{UAV}s and also possess challenges such as reliability and low-latency communication besides interference during aerial-ground communications. 
Solving these challenges is also essential for efficient control of \ac{UAVs} in \ac{5G} and beyond communications. It is noted that more flexible \ac{3D} beamforming will be deployed in the forthcoming \ac{5G} systems, to enhance both beamforming gain and interference mitigation by exploiting the angle resolutions in both azimuth and elevation dimensions of \ac{UAVs} in the sky\cite{cellular_UAV_4}. \ac{UAVs} for aerial-terrestrial interference management has been extensively studied in \ac{BS}-\ac{UAV} communication scenario over recent years\cite{uav_interference_1, uav_interference_2}. Hence, \ac{mmWave} beam alignment is critical in supporting the flying \ac{UAV}s with high mobility and semi-autonomous operations.

Fast \ac{mmWave} beam alignment can enhance the data throughput for both \ac{UAV}-\ac{UAV} and \ac{BS}-\ac{UAV} communications under \ac{5G} and beyond wireless systems. Especially, the availability of \ac{UAV} position information at lower frequencies (following the \ac{3GPP} standard \cite{UAV_3gpp_1}) may also provide scope for reliable communication in addition to increasing throughput. Position information for fast beam alignment has been recently studied under vehicular context in \ac{mmWave} systems \cite{position_beamalign_2, position_beamalign_3}. On the other hand, high mobility and autonomous operation of \ac{UAVs} will require frequent beam realignment as well. Therefore, a faster and reliable beam alignment using \ac{UAV} position information is crucial in enabling high data rate for \ac{mmWave} \ac{UAVs}. 

Existing works~\cite{channel_tracking1, channel_tracking3} proposed beam tracking schemes using Kalman filters with high processing complexity. However, these schemes are vulnerable to abrupt changes in environment when \ac{UAV}s are moving at high speeds and tracking error accumulates over time. On the other hand, conventional beam sweeping solutions \cite{Exh_beam_search} often have large overhead, which is also unacceptable with respect to high speed and autonomous movement of \ac{UAVs} as it requires frequent beam re-alignment. An alternate approach to this is to perform fast beam alignment for every change in \ac{UAV} position along the \ac{BS} coverage area. 
Existing works in vehicular environment proposed different beam alignment approaches for terrestrial systems based on black box function optimization\cite{blackbox_optim_1, blackbox_optim_2} and the use of contextual information \cite{position_beamalign_3, beam_learning_1}. 

Contextual information generally involves data from the sensors such as position information, antenna configurations, \ac{RSS} power etc. using low frequency control signals from \ac{3GPP}\cite{5gprotocol_book} whenever needed, as this information is used abundantly to reduce the beam training overhead. 
The authors in \cite{beam_learning_1} assumed position information as context and perform beam training over selected subset of beam-pairs by updating a database of channel strength information in an online manner. The proposed method suits well to the vehicular communication context with rapid channel variations in the environment. \ac{BS}-\ac{UAV} environment generally involves large coverage areas and high speed mobility of \ac{UAVs} resulting in frequent change in \ac{UAV} position. As a result, pre-processing of beam pairs and maintaining a database could be complex, non-generic and also cause significant overhead. In addition, learning a subset of beam-pairs based on the past beam measurements might not be essential as there are relatively less channel variations under this environment.
Black box optimization frameworks in \cite{position_beamalign_1, v2x_ml_1}, investigate application of supervised \ac{ML} techniques to beam alignment problem. These frameworks are computationally efficient but assumed a separate training data collection phase for proposed supervised learning environments.

In our work, we propose a generic \ac{DQN}-based framework using both black box and user-context optimization techniques, in an attempt to progress from the previous works for \ac{UAV} beam alignment problem. 
We approximate the beam alignment optimization problem for a multi-location environment using \ac{NN} techniques and learn best beam-pairs for any \ac{UAV} in these locations from their past beam measurements. This approximation also helps in using prior knowledge of the propagation environment to significantly reduce the beam-alignment overhead. Our algorithm takes into account \ac{UAV} location and finds an optimal beam pair to maximize the beamforming gain between \ac{BS} and \ac{UAV}. To identify the significant change in UAV location, we consider a grid environment with \ac{UAV} position information of the grid element as the user-context information. The main contribution of this paper is that we propose the \ac{RL}-based \ac{DQN} framework to optimize beam alignment between \ac{BS} and \ac{UAV} for any grid position inside the \ac{BS} coverage area, during uplink \ac{mmWave} communication. We benchmark our results against \ac{MAB}, exhaustive based approaches, and analyse against different coverage area requirements under ideal \ac{UMa}-\ac{nLoS} conditions. We also simulate the better training performance of \ac{DQN} framework generic to any \ac{UAV} position in an online manner under varying channel conditions.

The rest of the paper is organized as follows. Section~\ref{section:system_communication_model} presents the problem formulation and communication modelling, considered in this problem. 
Section~\ref{section:learning_formulation} discuss in detail about the proposed \ac{DQN} based \ac{RL} approach for beam alignment. 
Section~\ref{section:simulation_results} presents the comparison of the proposed \ac{DQN}-based \ac{RL} approach against \ac{MAB}-based learning and different coverage area requirements under ideal channel conditions. The section also discuss about the online learning of \ac{DQN} and its analysis under varying channel conditions. Section~\ref{section:conclusion_future_scope} summarizes the conclusion and future work. 

\vspace{-2mm}
\section{System and Communication Model}\label{section:system_communication_model}
In this section, we describe the system model, communication model for the learning framework using \ac{3GPP} protocol standards and also formulate the parameters to be used in the proposed learning framework. The objective of this problem is to maximize the beamforming between \ac{BS} and the \ac{UAV} to provide efficient communication under the defined environment and channel conditions.
%
\begin{figure}[h]
    \centering
    \captionsetup{justification=centering}
    \includegraphics[trim={0.95\columnwidth} 7cm {0.1\columnwidth} 3.5cm, clip, scale=0.65]{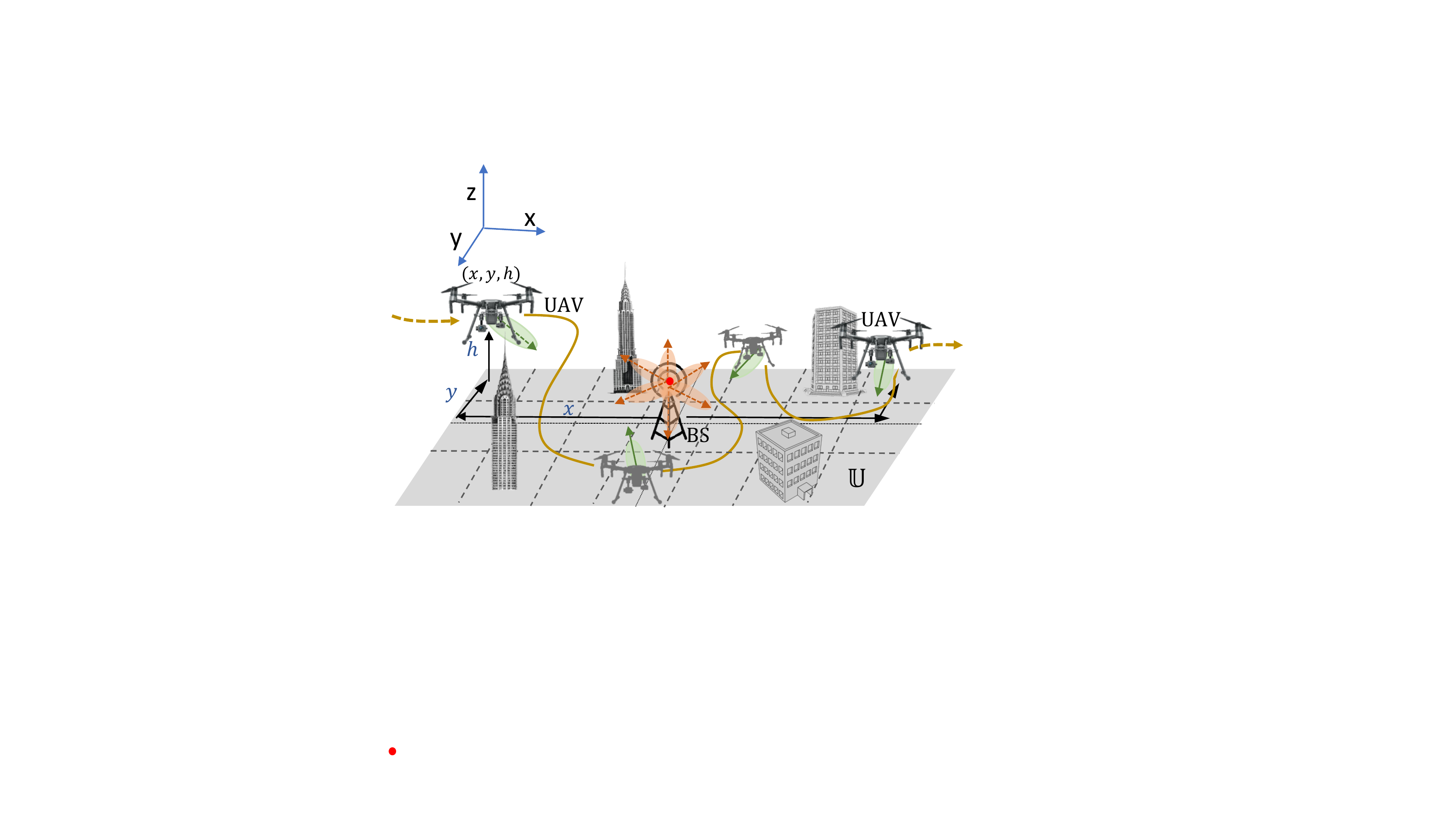}
    \caption{System Model}
    \label{fig:problem_formulation}
\end{figure}
%
%
    
    

\vspace{-0.65cm}
\subsection{System Model}\label{subsection:system_model}
We consider a cellular \ac{mmWave} \ac{MIMO} uplink communications between a single \ac{UAV} and a single \ac{BS} under \ac{UMa} environments. The \ac{BS} is fixed at \(\mathcal{O}(0,0)\) and communicates with the moving \ac{UAV} (hereafter used as \ac{UE}) using a multi-path \ac{mmWave} beamforming as shown in Figure~\ref{fig:problem_formulation}. 
The multi-antenna \ac{UE} hovers randomly and communicates with the multi-antenna \ac{BS} in the urban environment following \ac{5G} \ac{NR} standard protocol\cite{5gprotocol_book}. We consider an analog beamforming equipped with one \ac{RF} chain and \ac{ULA} structures of \(N_t\) and \(N_r\) antennas for both \ac{BS} and \ac{UE}, respectively. The \ac{UE} transmits (TX) while the \ac{BS} receives (RX) a radio signal in multiple beam directions following \(\mathcal{B_{\text{TX}}}\) and \(\mathcal{B_{\text{RX}}}\) codebook, respectively with angles defined as
\begin{equation} \label{eq:codebook}
 b_i = (i-1)\frac{\pi}{N}, i\in\{1,2,...,N\},
\end{equation}
where \(b_i\) represents a RF radio beam direction with a fixed narrow beam width (\(\frac{\pi}{N}\)), \(N\) represents \(N_t\), \(N_r\) antennas for \(\mathcal{B_{\text{TX}}}\) and \(\mathcal{B_{\text{RX}}}\) codebook, respectively. 
The codebook values are defined using the beamforming vectors \(\mathbf{w}_{\text{TX}}\) and \(\mathbf{w}_{\text{RX}}\) for \ac{UE} and \ac{BS}, respectively, given by \(\mathbf{w}(b_i)]_{n=0}^{N-1} =  \frac{1}{\sqrt{N}} \exp(j \frac{2\pi n d}{\lambda} \sin(b_i))\), 
%
where  \(b_i\in \mathcal{B}\), \(N=N_t\), \(\mathcal{B}=\mathcal{B_{\text{TX}}}\) and \(N=N_r\), \(\mathcal{B}=\mathcal{B_{\text{RX}}}\) for \(\mathbf{w}=\mathbf{w}_{\text{TX}}\) and \(\mathbf{w}=\mathbf{w}_{\text{RX}}\), respectively. Here, \(d=\frac{\lambda}{2}\) is the antenna spacing, \(\lambda\) is the wavelength and \(b_i\) is the \(i^{th}\) codebook direction \eqref{eq:codebook}. The \ac{UE} moves randomly along the \ac{3D} coverage area which is divided into multiple grids of equal size, the set enclosing them is denoted as \(\mathbb{U}\). The received signal measurement can be observed at the \ac{BS} for different TX-RX beam direction pairs and their timing information can be estimated using \ac{SS} block symbols, during beam selection procedure of the \ac{3GPP} beam-alignment protocol\cite{5gprotocol_book}. 
For simplicity, we define the time consumed by each such symbol as the \ac{TTU}, to measure the communication overhead of the learning procedure.
\vspace{-0.35cm}
\subsection{Communication Model}
\label{subsection:Communication_Model}
For the communication model, we consider a multi-path link (\ac{LoS} or \ac{nLoS}) radio channel between \ac{UE} at time \(t\) and \ac{BS} location \(\mathcal{O} \in \mathbb{R}^3\). 
\(\mathrm{UE}_l(t)\) represents the \ac{UE} position on grid index \(l\in\{0,1,....|\mathbb{U}|\}\) in the \ac{BS} coverage area \(\mathbb{U}\) at any time instant \(t\), given by 
\begin{equation}\label{eq:UEt}
    \mathrm{UE}_l(t)=(x_t, y_t, z_t),
\end{equation}
where \(\{x_t, y_t, z_t\}\in \mathbb{U}\). 
We assume \(\mathrm{UE}_l\) (with respect to \ac{BS} locations) is known during each \(P_1\) procedure of \ac{3GPP} beam access protocol. 
Let \(\theta_{\mathrm{tx,m}}\), \(\theta_{\mathrm{rx,m}}\) be the \ac{AoD} and \ac{AoA} of \(m^{\mathrm{th}}\) communication link  between \ac{BS} and \ac{UE}, respectively. 
The \ac{UE} transmits radio signals in a codebook direction from \(\mathcal{B}_{\text{TX}}\) while the \ac{BS} receives the signal through one of its multiple beam directions following \(\mathcal{B}_{\text{RX}}\) \eqref{eq:codebook}. Baseband equivalent of the received signal is given by
\begin{equation}
\begin{multlined}
\label{eq:recv_signal}
     y[k] = \underbrace{\sum_{m=0}^{M}\sqrt{P_{tx}}\beta_{m}\ \mathbf{w}_{\text{RX}}^H \mathbf{a}_R(\theta_{\mathrm{rx,m}}) \mathbf{a}_T^H(\theta_{\mathrm{tx,m}}) \mathbf{w}_{\text{TX}}x[k]}_\text{\(r[k]\)} + \nu[k],
\end{multlined}
\end{equation}
where \(P_{tx}\) is transmission power, \(M\) is the number of multi-paths or reflection points in the \ac{UMa} environment \cite{3GPP_avchannel_model}, \(\beta_m\), \(\mathbf{a}_R({\theta}_{\mathrm{rx,m}}) \in \mathbb{C}^{N_r}\), \(\mathbf{a}_T({\theta}_{\mathrm{tx,m}}) \in \mathbb{C}^{N_t}\) are the antenna channel gain and array response vectors for \(\theta_{\mathrm{rx,m}}\) and \(\theta_{\mathrm{tx,m}}\) along the $m^{th}$ communication link, respectively. Here, \(\mathbf{a}(\theta)]_{l=0}^{N-1} =  \frac{1}{\sqrt{N}} \exp(j \frac{2\pi l d}{\lambda} \sin(\theta))\), where \(\theta=\theta_{\mathrm{rx,m}}, N=N_r\) and \(\theta=\theta_{\mathrm{tx,m}}, N=N_t\) for \(\mathbf{a}_R({\theta}_{\mathrm{rx,m}})\) and \(\mathbf{a}_T({\theta}_{\mathrm{tx,m}})\), respectively. \(\mathbf{w}_{\text{RX}} \in \mathbb{C}^{N_r}\), \(\mathbf{w}_{\text{TX}} \in \mathbb{C}^{N_t}\) are the transmit and receive  unit-norm beamforming vectors, \(\nu[k] \sim \mathcal{CN}(0,W N_0)\) is the effective noise with zero mean and two-sided power spectral density \(\frac{N_0}{2}\), \(x[k]\) represents one \ac{OFDM} symbol of the time-domain transmitted signal with bandwidth \(W\) and  \ac{TTU} time period with \(\frac{1}{K} \sum_{k=0}^{\text{K}}\|x[k]\|^2 = 1\). Here, \(k=0,1,...K\) is the number of samples spanned over \ac{TTU} time. In this work, we assume the channel measurements \(\mathbf{H}_m\) of the \(m^{th}\) multi-path link \(\mathbf{H}_m(\theta_{\text{tx},m},\theta_{\text{rx},m})\triangleq \beta_{m} \mathbf{a}_R({\theta}_{\mathrm{rx,m}}) \mathbf{a}_T^H({\theta}_{\mathrm{tx,m}})\) to follow \ac{3GPP} \ac{UMa} channel conditions\cite{3GPP_avchannel_model}. We define \(r[k] = \sum_{m=0}^{M}\sqrt{P_{tx}} \mathbf{w}_{\text{RX}}^H \mathbf{H}_m(\theta_{\text{tx},m},\theta_{\text{rx},m}) \mathbf{w}_{\text{TX}} x[k]\). Then, the \ac{SNR} is given as \(\mathrm{SNR} = \frac{\frac{1}{K}\sum_{k=0}^{K}\norm{r[k]}^2}{N_0 W}\) and overall rate measurement \(R\) in bits per channel use  is given by
\begin{equation}\label{eq:rate}
R = \log(1+\mathrm{SNR}).
\end{equation}
Thus, the beam alignment for the above formulation is realized through data rate measurements. The optimal beam-pair for \ac{UE}-\ac{BS} is selected based on their data rates under the scenario mentioned in Section~\ref{subsection:system_model}.  
\subsection{Problem Formulation}
We consider an uplink communication between \ac{BS} and \ac{UE} following \ac{3GPP} beam access protocol \cite{3GPP_protocol, 5gprotocol_book}. 
The beamforming protocol in general involves mainly three procedures, Initial communication (used as \(P_1\) procedure), beam selection (used as \(P_2\) procedure) and beam refinement (used as \(P_3\) procedure) \cite{3GPP_protocol}. 
Here, we assume \ac{UE} and \ac{BS} share position information and sequence of beam-pairs with one another at low frequencies during \(P_1\), respectively and sweep the selected sequence of beam pairs during $P_2$ following \ac{5G} communication~\cite{5gprotocol_book}. 
We formulate the \ac{3GPP} based beam-pair alignment learning through \(P_1\) and \(P_2\) procedures, and maximize the beamforming gain for any \ac{UE} position around the \ac{BS} coverage area \(\mathbb{U}\). Timing information of these procedures assumes \ac{3GPP} frame structure and time for each \ac{OFDM} slot is defined in our work as the \ac{TTU}\cite{3GPP_protocol}. We consider \ac{RSS} of radio signal and radio beam pair directions (both TX and RX) as the known and unknown parameters of this multi-location environment, respectively. We define the state and action spaces for learning based beam-pair alignment as follows:
\begin{equation}\label{eq:env_formulation}
\begin{aligned}
(\mathcal{E}): \
\mathcal{S} &= \{\mathrm{UE}_l, 1\leq l \leq|\mathbb{U}|\} \\
\mathcal{A} &= \{(b_p, b_q), 1\leq p \leq|\mathcal{B}_{\mathrm{TX}}|, 1\leq q \leq|\mathcal{B}_{\mathrm{RX}}|\}, \\
\end{aligned}
\end{equation}
where \(\mathrm{UE}_l\) is the location of \ac{UE} within coverage area \(\mathbb{U}\)~\eqref{eq:UEt} while \(\mathcal{B}_{\mathrm{TX}}\), \(\mathcal{B}_{\mathrm{RX}}\) is the beam codebook sets at \ac{UE} and \ac{BS} side respectively~\eqref{eq:codebook}. 

\ac{RL} is an interactive learning problem following a \ac{MDP}. In this work, the \ac{RL}-based beam alignment problem is modelled as a \ac{POMDP}. At any time instant \(t\), we define the parameters \(s_t=\{(s',a') \ s'\in \mathcal{S},a' \in \mathcal{A}\}\), \(a_t \in \mathcal{A}\) and \(r_t \in \mathcal{R}\) where \(s_t\), \(a_t\), \(r_t\) are the state, action and reward at time instant \(t\). 
\(a'\) corresponds to the set of previous actions applied for state transitions until the time instant \(t\). Data rate measurements computed on applying each action are considered as the rewards for the problem. We denote \(o_t=\{a_{t-1}, s_{t-1}, a_{t-2}, s_{t-2}, ...., a_{1}, s_{1}\}\) as the observed history of all such state information and past actions. After the \ac{3GPP} initial communication procedure with \ac{UE}, \ac{BS} starts with a random receiving beam direction and then proceeds towards the maximum beamforming gain by applying actions and undergoing state transitions, accordingly. The current applied action becomes part of the next state, undergoing state transition. The objective of this problem can be formulated as 
\begin{equation}\label{eq:rl_obj_fn}
\begin{aligned}
(P1)&: \max_{\{\pi(a_t|o_t)\}} \sum_{i=t}^{\infty} \gamma^{i-t}\mathbb{E}_{\pi}[r_{a_i}(i)],\\
s.t. &\\ r_{a_t}(t) &= 
    \begin{cases}
        1 & \text{if } R(a_t) \geq R_{max}(o_t)\\
        -1              & \text{otherwise}
    \end{cases},\\
    &\gamma \in (0,1],\\
\end{aligned}
\end{equation}
where \(R_{max}(o_t)\) is the optimal data rate measurement observed among the information history \(o_t\) until time instant \(t\), \(r_{a_t}(t)\) and \(R(a_t)\) are the rewards and data rate measurement observed on applying action beam-pair $a_t$, respectively. We follow \ac{DQN} approach to solve this \ac{RL} objective problem. 
%
\section{Implementation}\label{section:learning_formulation}
\ac{DQN} is a value-based approach used generally in the context of \ac{RL} \cite{Mnih_2015}. The approach learns an optimal approximated policy of states mapping to actions \(\pi(s) = a\) by parameterizing and estimating state-action value function \(Q(s,a;\theta)\) using \ac{DNN}. 
We denote the primary \ac{DNN} network weight matrix and target \ac{DNN} network weight matrix as \(\theta\) and \(\overline{\theta}\), respectively \cite{Mnih_2015}. We consider a fully connected \ac{DNN} for both the networks where \(\overline{\theta}\) is updated with primary network parameters \(\theta\), after every \(K\) iterations. The input of \ac{DNN} is given by the variables in \(s_t\). The intermediate layers are fully connected linear units with \ac{ReLU} by using the function \(f(x)= \max{(0,x)}\) and the output layer is composed of linear units, which are in one-one corresponding relationship with the action space \(\mathcal{A}\). We consider initialization of bias and weights of these layers with zeros and Kaiming normalization, respectively.
\begin{algorithm}
\caption{DQN approach} \label{alg:dqn_rl_beamalignment}
\KwIn{The set of UAV x,y,z location coordinates and training episodes M}
Algorithm hyper-parameters: learning rate \(\xi \in (0,1]\), discount rate \(\gamma \in [0,1)\), \(\epsilon\)-greedy rate \(\epsilon \in (0,1]\), update steps \(K\);\\
Initialization of replay memory \(M\) to capacity \(C\), the primary Q-network with parameters \(\theta_1\), the target Q-network with parameters \(\theta_2\)\\
\(\mathcal{S},\mathcal{A}\): State and Action space of DQN agent \\
\For{episode \(\leftarrow 1\ to\ M\) \tcp{for each episode}}
{
    Initialization of \(s_1\) by executing a random action \(a_0\)\\
    n=0, \(N\rightarrow\)Episode limit\\
    \While{\(True\)}
    {
        \uIf{\(p_{\epsilon} < \epsilon\)} {select a random action \(a_t \in A\)}
        \Else{select \(a_t = {argmax}_{a\in \mathcal{A}} Q(s_t,a, \theta)\)}
        \ac{BS} applies \(a_t\) over the channel, receive signal for \((t+1)^{th} \) episode during uplink communication\\
        UE observes \(S^{t+1}\) and calculate the reward\\
        Store transition \(e = (s_t, a_t,r_{t+1},s_{t+1})\) in replay memory \(D\)\\
        Sample random minibatch of transitions \(U(D)\) \\
        Compute Loss and Perform gradient descent for \(Q(s,a;\theta)\)\\
        Update the target network parameters \(\theta_2 = \theta_1\) after every \(K\) steps\\
        \(n = n + 1\) \tcp{Increment episode time}
        \uIf{done or \((n = N)\)}{break \tcp{End episode}}
    }
}
\end{algorithm}

At a time instant \(t\), \ac{DQN} selects action $a_t\in \mathcal{A}$ and perform forward propagation of \(Q(s_t,a_t;\theta)\) following \(\epsilon\)-greedy policy. A memory buffer of experiences \(D_t=\{e_1,e_2, e_3,...,e_t\}\), \(e_i = (s_i, a_i, r_{i+1}, s_{i+1})\) are collected, where a mini batch of them \(U(D)\) are randomly sampled and sent into \ac{DQN} \cite{Mnih_2015}. During back propagation, a \ac{MSE} loss function is computed between primary, target networks and \(\theta\) is updated using \ac{SGD} and Adam Optimizer as
\begin{equation}\label{eq:sgd}
\theta_{t+1} = \theta_{t} - \xi_{\mathrm{Adam}} \nabla \mathcal{L}^{\mathrm{DQN}}(\theta_{t}),
\end{equation}
where \(\xi_{\mathrm{Adam}}\) is the learning rate, \(\nabla \mathcal{L}(\theta_{t})\) is the gradient of the \ac{DQN} loss function. 
Complete steps followed by DQN for RL based beam alignment problem are shown in Algorithm~\ref{alg:dqn_rl_beamalignment}. 
Here, we define episode as the consecutive set of actions applied on the starting state until it reaches the terminal state with maximum beamforming gain for that location. In order to prevent episodes with infinite set of actions during training, we confine maximum episode length to exhaustive set of beam pairs possible under the chosen antenna configuration. 

\begin{figure}[!hpt]
    \centering
    \captionsetup{justification=centering}
    \includegraphics[
    trim=0.7cm .7cm 1.5cm 1.8cm,
    clip,
    width=\linewidth,
    keepaspectratio,
    scale=0.7,
    ]{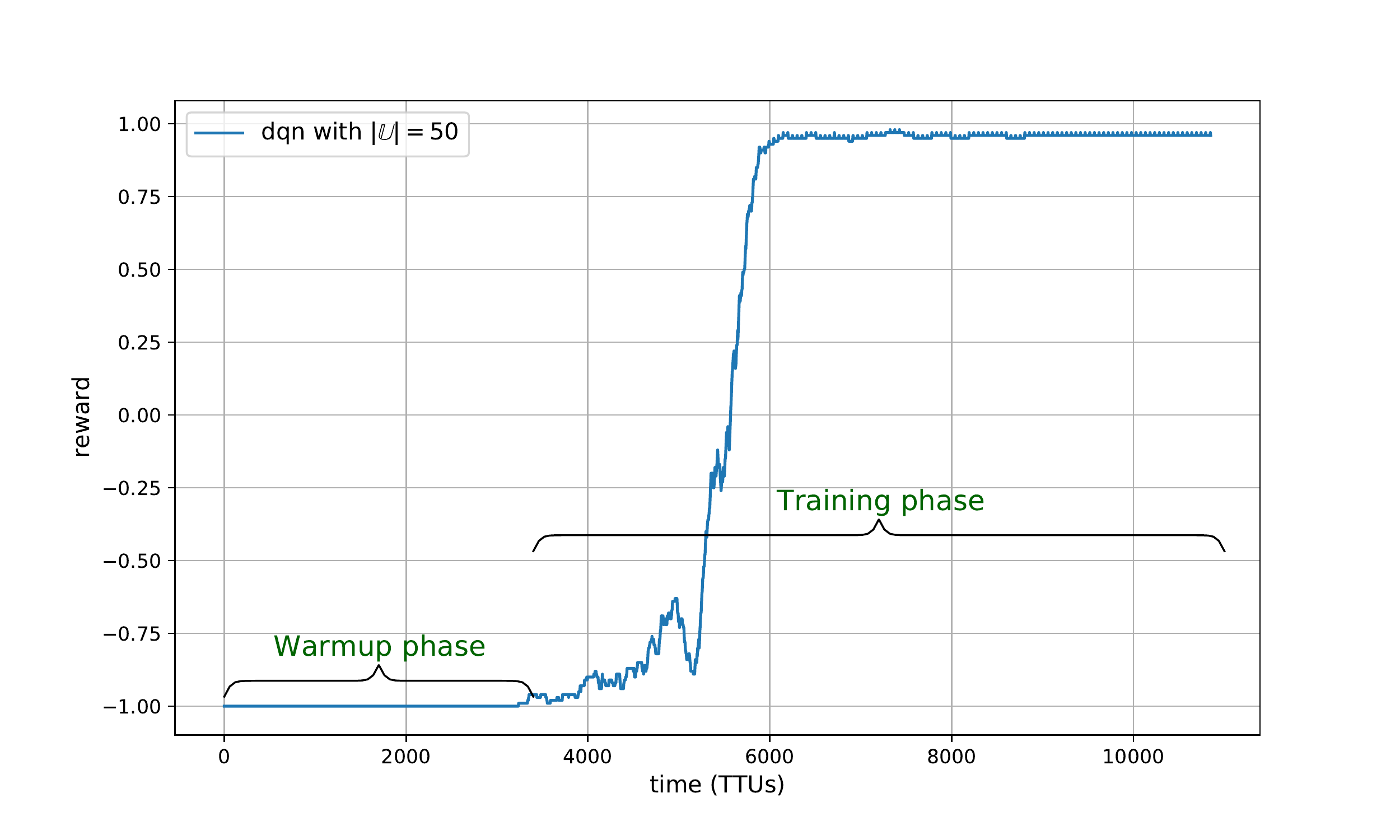}
    \caption{\ac{DQN} training procedure consisting of two phases namely, Warmup phase and Training phase}
    \label{fig:warmup_training_phases}
\end{figure}

As the \ac{RL} learning objective formulation involves both current data rate \(R_t\) and best observed data rate \(R_{\mathrm{max}}(o_t)\) measurements (shown in \eqref{eq:rl_obj_fn}), we consider the overall online training procedure of \ac{DQN} framework under two phases namely, Warmup phase and Training phase as shown in Fig.~\ref{fig:warmup_training_phases}. During the Warmup phase, the exploration is set to maximum, in order to observe the best possible data rate for the given \ac{UE} location by applying maximum episode length of actions. During the Training phase, the algorithm continues to reduce its exploration and move towards exploitation following \(\epsilon\)-greedy policy. The episode starts with initial action corresponding to $o_{t}$ and applies next actions to reach the terminal state as quickly as possible. The Warmup phase results in extra training time at the start but favours quick convergence of \ac{DQN} during training phase resulting in faster beam-alignment training for the multi-location environment.
\section{Simulation results}\label{section:simulation_results}
%
As described in Section~\ref{section:learning_formulation}, we implement the \ac{RL}-based beam alignment using \ac{DQN}, following \((P1)\) objective \eqref{eq:rl_obj_fn} and steps mentioned in Algorithm~\ref{alg:dqn_rl_beamalignment}. Similarly, we implement the \ac{MAB}-based approach for our problem (following existing works~\cite{beam_learning_1, mab_beamalign_1}) using \ac{gUCB} algorithm. 
Traditional approach mainly involves exhaustive search over entire action space \(\mathcal{A}\), to find the best beam pair with maximum beamforming gain between \ac{UE} and \ac{BS}. In our work, the exhaustive approach in general causes significant communication overhead (\(\mathcal{O}(|\mathcal{A}|)\)) with higher antenna elements as frequent beam scanning is required for every change in grid element unit of \ac{UE} inside \(\mathbb{U}\). On the other hand, \ac{RL} and \ac{MAB} learning-based methods once converged, can significantly reduce the communication overhead during $P_2$ procedure and maximize beamforming gain in \(\mathcal{O}(1)\) time. 

In this section, we first investigate the training performance of our proposed \ac{RL}-based approach against \ac{MAB}-based learning and different coverage area requirements in order to maximize beam alignment under ideal \ac{UMa}-\ac{nLoS} channel conditions. Later, we combine these observations and perform online beam alignment in the presence of channel variation conditions.

\begin{figure}[!hpt]
    \centering
    \captionsetup{justification=centering}
    \includegraphics[
    trim=0.7cm .7cm 1.5cm 1.5cm,
    clip,
    width=\linewidth,
    keepaspectratio,
    scale=1.0,
    ]{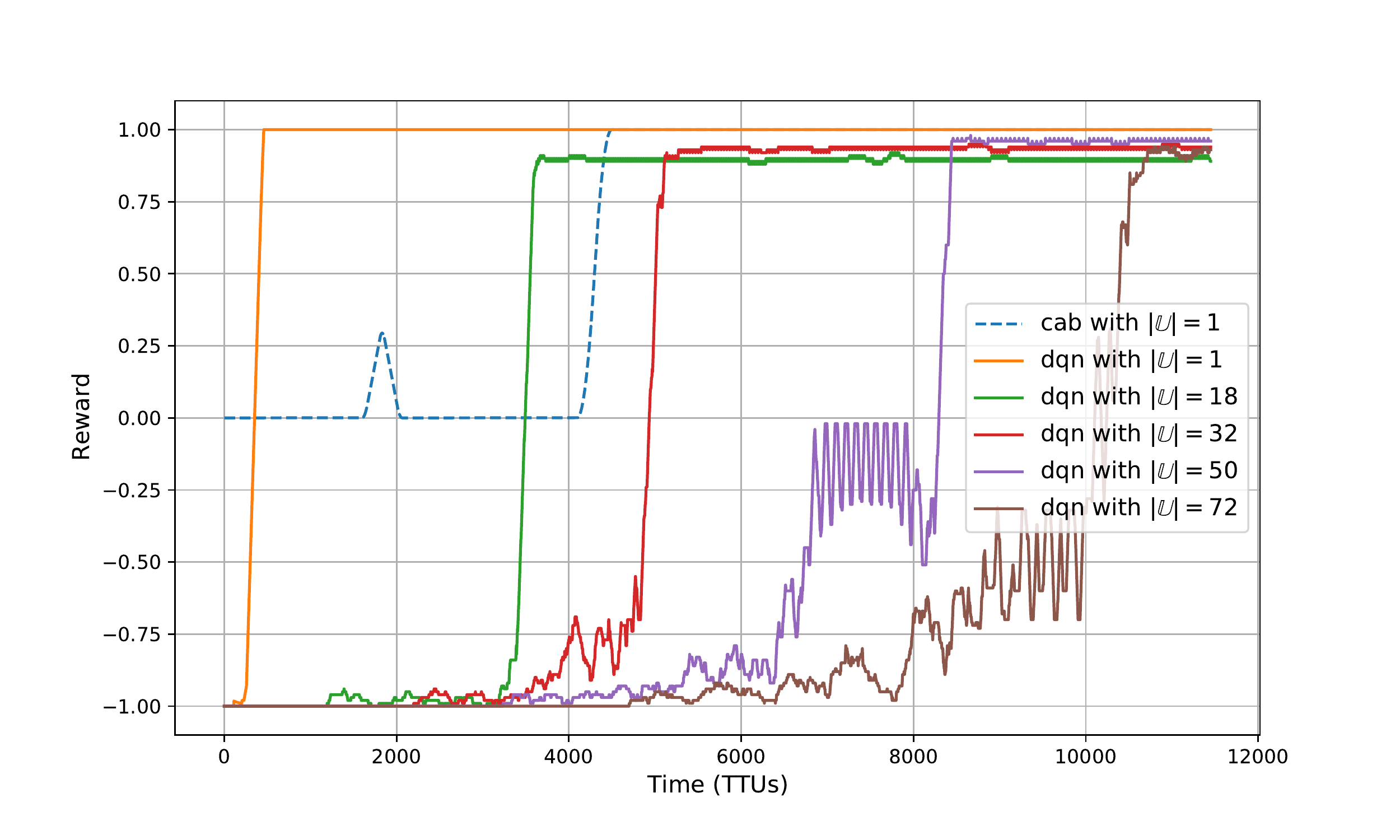}
    
    
%
 %
    \caption{\ac{DQN} reward-time plots for various \ac{BS} coverage area requirements for ideal \ac{UMa}-\ac{nLoS} channel conditions}
    \label{fig:dqn_covarea_rwdplot_2}
\end{figure}
\vspace{-.75cm}
\subsection{DQN training performance}\label{subsection:dqntraining_performance}
As shown in Fig.~\ref{fig:dqn_covarea_rwdplot_2}, brown and blue plots represent the accumulated rewards over time for \ac{DQN} and \ac{gUCB}, respectively, under single \ac{UE} grid location ($\mathbb{U}=1$) and ideal \ac{UMa}-\ac{nLoS} channel conditions with 6 reflection points. 
Our results show that the \ac{DQN}-based approach quickly accumulate the rewards over \ac{TTU} time for learning the optimal beam pair. The \ac{gUCB} approach involves iterating over exhaustive action space $\mathcal{A}$, thus consuming more time at every episode of its training procedure. On the other hand, \ac{DQN} uses its warmup phase to quickly determine the best possible data rate measurement for the current state and then learn the determined best rate during training phase, by applying actions in a iterative manner. With increase in \ac{BS} coverage area, 
the \ac{gUCB}-based approach is observed to be unreliable for multiple grid locations without frequent re-training. This is due to substantial change in \ac{RSS} measurements w.r.t (TX,RX) beam pairs for every \ac{UE} grid location.

On the other hand, \ac{DQN} agent with the same architecture is very much reliable to learn beam alignment across multiple \ac{UE} locations for different coverage area requirements under the same ideal channel conditions as shown in Figure~\ref{fig:dqn_covarea_rwdplot_2}. 
After the convergence is obtained, \ac{BS} can quickly align using learnt optimal beams for any \ac{UE} position within the coverage area without any re-training. Also, the learning is observed to be relatively quicker in convergence with increase in coverage area of \ac{BS}. With increase in coverage area, neighbouring grid elements with similar optimal beam pairs converge \ac{DQN} faster as part of its \ac{MDP} process, resulting in average less number of training iterations per grid element alongside convergence. 
Thus, the \ac{DQN} agent can achieve faster and reliable beam alignment due to its training procedure and the neighbourhood grid element convergence.
\subsection{online DQN performance with channel variation and shadow fading}
\begin{figure}[!hpt]
    \begin{subfigure}{\columnwidth}
    \centering
    \captionsetup{justification=centering}
    \includegraphics[
    trim=0.7cm .7cm 1.5cm 1.8cm,
    clip,
    width=\columnwidth,
    scale=0.8,
    ]{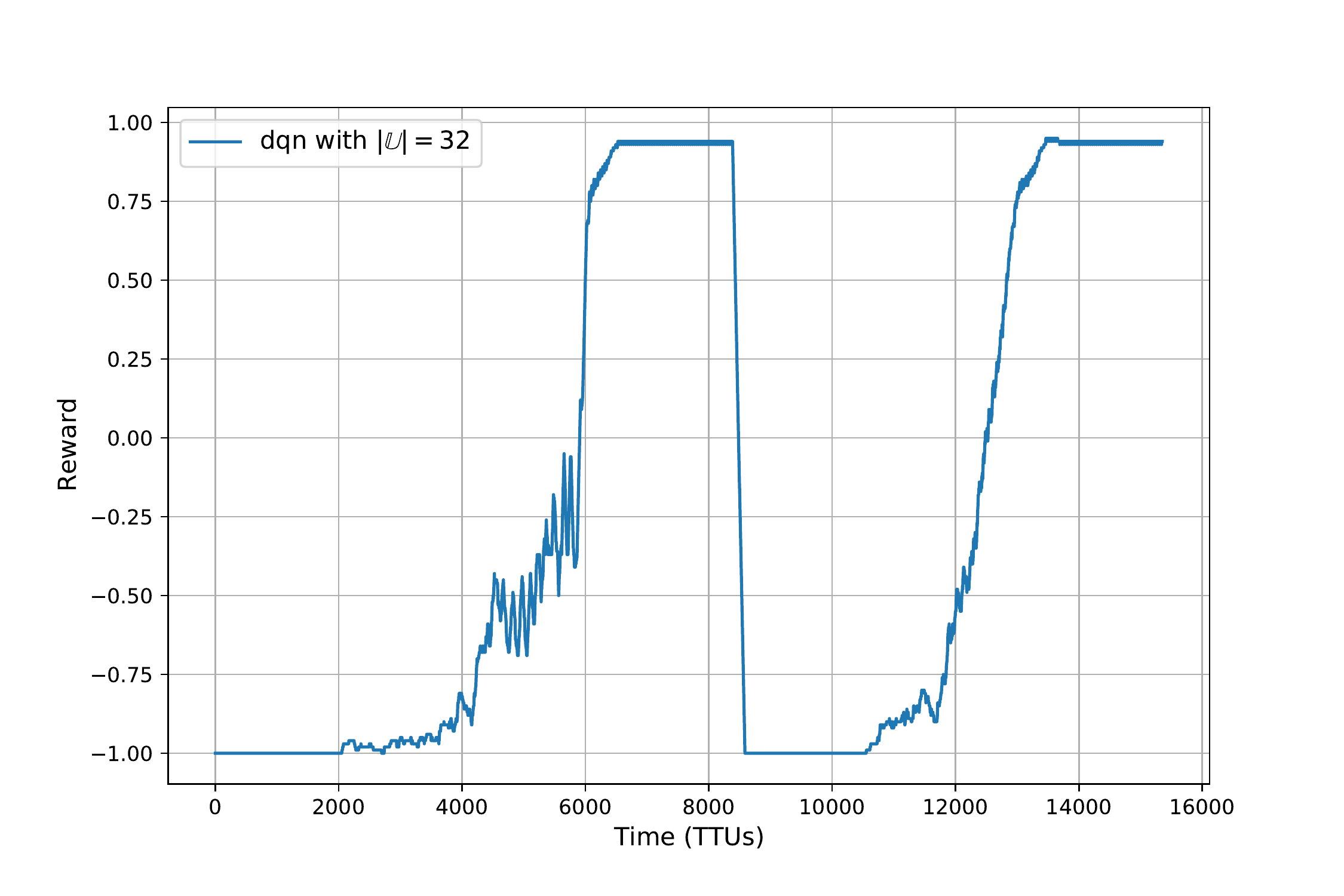}
    \caption{Reward plot for DQN based beamlearning with channel variation including shadow fading, thermal noise and rician fading}
    \label{fig:dqn_channelvariation_rwdplot}
    \end{subfigure}
    
    
    \begin{subfigure}{\columnwidth}
    \centering
    \captionsetup{justification=centering}
    \includegraphics[
    trim=0.75cm 0.75cm 1.5cm 0.75cm,
    clip,
    width=\columnwidth,
    keepaspectratio,
    scale=0.8,
    ]{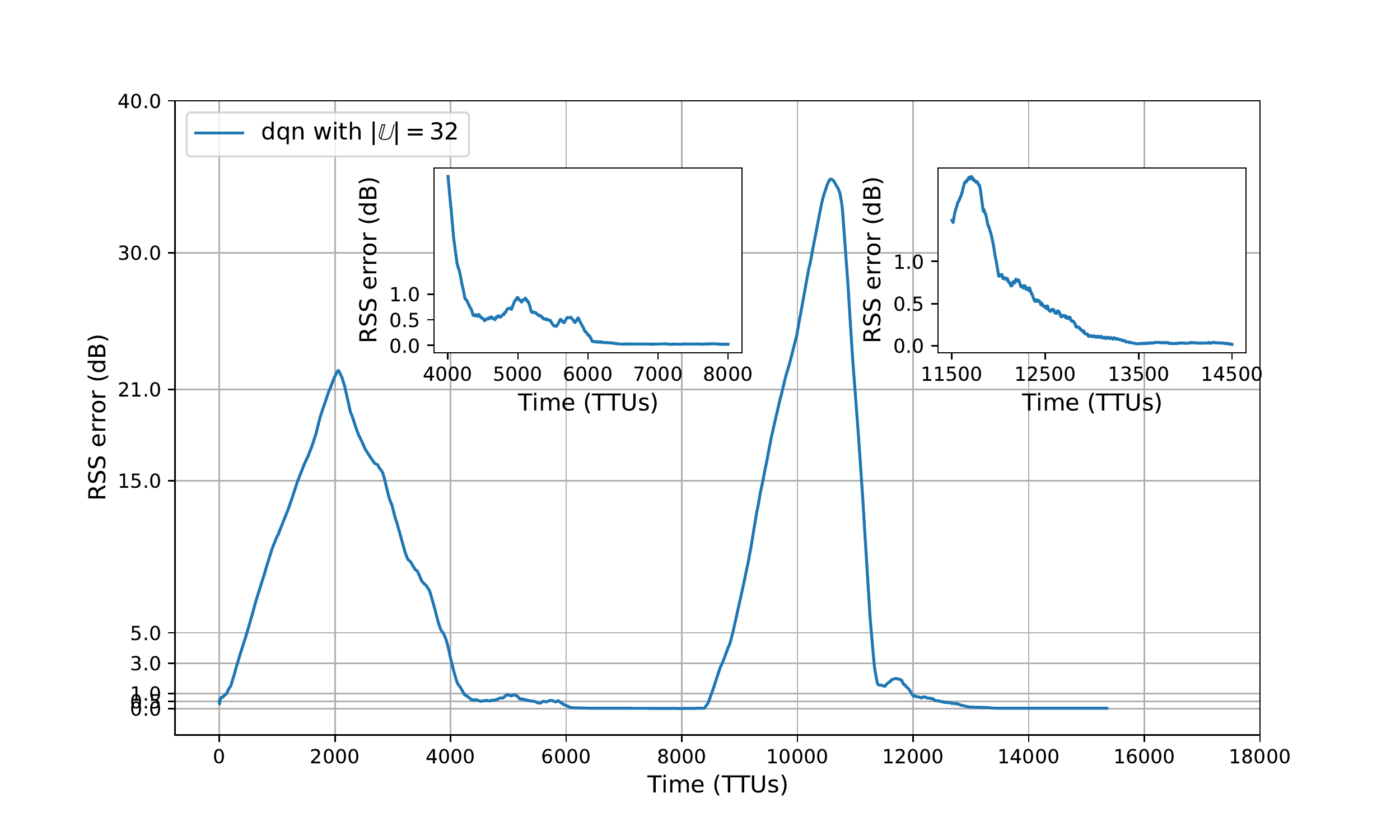}
    \caption{RSS Error plot for DQN based beam learning with channel variation including shadow fading, thermal noise and rician fading}
    \label{fig:dqn_channelvariation_errorplot}
    \end{subfigure}
    
    \caption{DQN-based beam learning performance under UMa-nLoS to UMa-LoS varying channel conditions}
    \label{fig:dqn_channelvariation_plots}
\end{figure}

In this subsection, we plot \ac{DQN} training performance in real-time conditions in an online manner by considering change in channel conditions, thermal noise, slow fading and slow channel variation at \ac{UE} grid locations as shown in Fig.~\ref{fig:dqn_channelvariation_plots}.

\begin{table}[h]
    \caption{Simulation Parameters}
    \label{tab:sim_parameters}
    \centering
    \begin{tabular}{|c|c|}
    \hline
    \rowcolor{cyan}
    Parameters & Value \\
    \hline
    \ac{mmWave} Channel & \ac{UMa} \\
    \ac{mmWave} freq & \(30\) GHz \\
    carrier spacing freq \(df\) & \(60\) kHz \\
    antenna element spacing \(d\)    & \(0.5\)\\
    Transmit power \(P_{\mathrm{tx}}\) & \(0\) dBm\\
    Transmit antenna elements \(N_{tx}\) & 8 \\
    Receiving antenna elements \(N_{rx}\) & 8 \\
    Noise Level \(N_0\) & -174 dBm \\
    BS location    & \([0,0,25]\) \\
    coverage xloc \(\mathbb{U}_{\mathrm{xloc}}\) & \([-60,60,20]\) m \\
    coverage yloc \(\mathbb{U}_{\mathrm{yloc}}\) & \([-60,60,20]\) m \\
    coverage zloc \(\mathbb{U}_{\mathrm{zloc}}\) & \([41.5,81.5]\) m \\
    Cardinality of State Space \(|\mathcal{S}|\) & \(72\) \\
    Cardinality of Action Space \(|\mathcal{A}|\) & \(64\) \\
    \hline
    \end{tabular}
\end{table}
Fig.~\ref{fig:dqn_channelvariation_rwdplot} and Fig.~\ref{fig:dqn_channelvariation_errorplot} plot the rewards and RSS error of the \ac{DQN} learning agent, respectively. We define RSS error as the expectation of the difference between \ac{DQN}-agent and exhaustive approach \ac{RSS} errors over \ac{BS} coverage area. For these simulations, we consider a (\(\mathbb{U}=32\)) coverage area environment between \ac{BS} and \ac{UE}. The environment is equipped with thermal noise, shadow fading, \ac{UMa}-\ac{nLoS} to \ac{UMa}-\ac{LoS} channel conditions along with a slow rician channel variation. The parameters used for this simulation are disclosed under Table.~\ref{tab:sim_parameters}. Depending on the defined coverage area, different path loss models including the aerial view of \ac{UE} for both UMa-nLoS and UMa-LoS are included following \ac{3GPP}\cite{3GPP_avchannel_model}. It is noted here that same set of hyper-parameters are used throughout the \ac{DQN} simulation. We observe that \ac{DQN} agent performs similar to previous results, converging well under both varying channel conditions. The disturbance in the smoothness of the reward plots observed in Fig.~\ref{fig:dqn_channelvariation_rwdplot} could be due to the impact of channel variation under \ac{nLoS} conditions. From Fig.~\ref{fig:dqn_channelvariation_errorplot}, we observe that the RSS error increases at first during online training phase and then decreases to zero on model convergence under both varying \ac{LoS} and \ac{nLoS} conditions. This shows that \ac{DQN}-framework on convergence, achieves optimal data rate measurements similar to that of exhaustive approach.   
\vspace{-1.25cm}
\section{Acknowledgements}
The research was supported by 6G Flagship programme, Finland and the Engineering and Physical Research Council (EPSRC), U.K., under Grant EP/W004348/1.
\section{Conclusion and Future Work}\label{section:conclusion_future_scope}
In this paper, we developed a learning-based beam alignment framework for \ac{mmWave} \ac{MIMO} uplink \ac{BS}-\ac{UAV} communication. We proposed a \ac{RL}-based framework using \ac{DQN} to maximize the beam alignment for any \ac{UAV} position within the \ac{BS} coverage area following \ac{3GPP} standard. 
%
%
We also analyze the same \ac{DQN} architecture over different coverage requirements under ideal channel conditions to demonstrate the generalization of the proposed \ac{RL}-based framework for beam learning. Our results show that, the proposed approach significantly outperforms the training performance of the \ac{MAB}-based method and also learns optimal beam pairs comparable to that of the heuristic method in an online manner under varying channel conditions.
Having shown some promising results, we will address the full capabilities of these generic learning architectures under higher \ac{MIMO} antenna configurations, large number of beam-directional pairs, interference mitigation etc. in future. 
\ifCLASSOPTIONcaptionsoff
  \newpage
\fi
%
%
%
%
%
\vspace{-0.3cm}
\end{document}